# Title: On the physical basis of biological signaling by interface pulses.

**Authors:** B. Fichtl[1,2], I. Silman[4], M. F. Schneider[3*]

[1]*University of Augsburg, Experimental Physics I, Augsburg, 86159, Germany.*

[2]*Nanosystems Initiative Munich NIM, Schellingstr. 4, 80799 München, Germany.*

[3]*Medizinische und Biologische Physik, Technische Universität Dortmund, Otto-Hahn Str. 4, 44227 Dortmund, Germany.*

[4] *Department of Neurobiology, Weizmann Institute of Science, Rehovot 76100, Israel*

\* Corresponding author

**Abstract**

Currently, biological signaling is envisaged as a combination of activation and movement, triggered by local molecular interactions and molecular diffusion, respectively. However, we here suggest, that other fundamental physical mechanisms might play an at least equally important role. We have recently shown that lipid interfaces permit the excitation and propagation of sound pulses. Here we demonstrate that these reversible perturbations can control the activity of membrane-embedded enzymes without a requirement for molecular transport. They can thus facilitate rapid communication between distant biological entities at the speed of sound, which is here of the order of 1 m/s within the membrane. The mechanism described provides a new physical framework for biological signaling that is fundamentally different from the molecular approach that currently dominates the textbooks.

Starting in the late 1970s, Kaufmann developed a theory deriving biological processes by starting from the 2$^{nd}$ law of thermodynamics[1–5]. To this end he followed Einstein's approach to thermodynamics [6–8], and applied it to the interface, which is ubiquitous in biology and must obey the 2$^{nd}$ law. Our present study is based on, inspired, and critically shaped by Kaufmann's work, in particular by his theories on action potentials[2] and the origin of catalysis [9,10].

How cells communicate and interact, how the numerous localized processes within a cell are integrated and how multicellular life arises in principle, are two of the most seminal issues in cell biology. Currently, intra- and intercellular signaling are described as cascades of binding and molecular activation events, with diffusion providing the physical mechanism for movement of signaling molecules to their targets. An important example, recognized by the awards of the Nobel Prizes for Physiology and Chemistry in 1994 and 2012, respectively, is the outcome of the extracellular binding of a hormone or neurotransmitter to a G-Protein coupled receptor (GPCR). The binding event evokes activation of G proteins at the intracellular surface, resulting in generation of second messengers, such as cyclic AMP (cAMP). The second messenger, in turn, may stimulate downstream functions, such as the degradation of glycogen to glucose [11]. The underlying physical principle is believed to be a conformational change induced in the receptor protein by binding of a ligand results in catalytic processes and/or disassociation of protein complexes, followed by diffusion of either low molecular weight species or protein subunits to their sites of action [12]. We suggest that a completely different mechanism of signal transduction may also function, in which propagating pulses serve as the means of communication.

A interface continuum, such as the cell membrane, constantly experiences a variety of external perturbations. These include pressure and temperature variations, changes in ionic strength or pH that are often a consequence of catalytic reactions, and binding or release of proteins and low molecular weight ligands. If the membrane is sufficiently decoupled from the bulk over the timescale of the perturbation, the perturbation will begin to propagate along the membrane. Importantly, this is not a hypothesis, but is inevitable, following directly from the second law of thermodynamics. (Footnote: If propagation is impaired, but dissipation still suppressed, a perturbation may lead to local oscillations, which we will not discuss here, since we consider the membrane as a continuous system, and hence assume conservation of momentum and propagation). As already mentioned at the beginning, this conceptual framework originates from the theoretical work of Konrad Kaufmann [2].

In recent years we were indeed able to demonstrate experimentally that lipid interfaces are capable of efficiently supporting the propagation of both linear and non-linear acoustic pulses [13–19]. These pulses propagate with velocities of the order of 0.1–1m/s, where the velocity follows the compressibility of the interface, even though – somewhat counter intuitively – the viscosity of the surrounding media seems to influence the velocity as well[20,21]. A minimum in velocity near the maximum in isothermal compressibility was observed, and identified as a typical fingerprint for sound propagation. Since sound transmission is an adiabatic process, we concluded that little or no heat leaves the pulse during propagation, presumably because the timescale of oscillation is shorter than the time required for significant heat transfer from the interface to the bulk. Another reason that the energy is retained within the interface, thus permitting propagation over macroscopic distances, is the large difference in compressibility between the membrane and the aqueous phase, which implies poor impedance coupling similar to that involved in wave guiding [13].

These acoustic waves can be excited by a variety of variables. Thus, optical, mechanical and chemical (pH) excitation have been demonstrated [14,19,22]. Interestingly, these variables not only allow evocation of pulses, but are, themselves, directly affected by the propagating pulses: thus, the local pH at the monolayer surface changes in phase with fluorescence emission, surface potential and surface pressure [15,16,19]. Hence, such interfacial pulses are not merely a mechanical phenomenon, but correspond to a reversible thermodynamic state change of *all* thermodynamic variables. This, in turn, opens up an entirely new range of possibilities for intra- and intercellular cell communication.

Needless to say, communication by sound clearly "beats" diffusion in all respects: it is adiabatic, directed, *ca*. $10^6$-fold faster, and scales linearly with distance, thus functioning well not only over small but also over large distances. Moreover, communication does not require large amounts of material to be transported. The experiments presented here provide clear evidence how pulses, propagating within a lipid membrane, can regulate the catalytic activity of a distant, embedded enzyme. They thus provide strong support for a role for adiabatic propagation in biological communication and signaling.

Our system – a simple mimic of a cell membrane - is based on an enzyme embedded in an excitable lipid monolayer. The setup allows local detection of the mechanical component of a two-dimensional propagating pulse simultaneously with enzymatic activity, and hence gives direct access to pulse-enzyme interactions (Fig. 1). As we recently demonstrated, a "puff" of acidic gas evokes the excitation of a propagating pulse [19]. Our observation that the monolayer can be stimulated by either hydrogen chloride, acetic acid or carbon dioxide clearly disassociates the excitation process from the respective Brønsted acids. Furthermore, the excitation strength depends strongly on the $pK_a$ of the lipid monolayer, thus pinpointing the origin of the phenomenon: protonation of the lipid head groups due to local acidification. The velocity of the pulses is determined by the state of the monolayer, represented by a particular compressibility. In the range of the phase transition the compressibility increases strongly, *i.e.*, the monolayer softens, resulting in a minimum in propagation speed. This was described earlier in greater detail, and demonstrates that the pulse propagation is indeed an adiabatic effect [14].

To test the hypothesis that these pulses (which originate in the same principles as sound) are capable of forming a new pillar in biological communication, we incorporated into the lipid monolayer the enzyme acetylcholinesterase (AChE), extracted and purified from the electric organ of *Torpedo californica*. The dimeric form employed here bears a covalently attached glycophosphatidylinositol (GPI) anchor, added post-translationally, at the C-terminus of each subunit [23–25]. The four fatty acid chains of the PI moieties are responsible for anchoring the enzyme to the membrane surface [26]. Enzymatic activity is measured by monitoring the cleavage of the chromophoric substrate, acetylthiocholine, at a wavelength of ~410 nm [27,28].

Figures 2A and B display the time-courses of simultaneous detection of enzymatic activity and lateral pressure during a travelling pulse elicited by HCl. The pulse propagates along the surface at a speed of ~0.5 m/s. Upon its arrival at the window of detection, where the enzymatic activity is recorded, substrate hydrolysis, and hence the catalytic rate, are clearly altered (Fig. 2A). After a strong, ~19-fold, increase, enzymatic activity appears to cease completely before reverting to its baseline value (Fig. 2B). While the increase in activity varied between 4-19-fold throughout our studies, the coupling between activity and pulse was observed in more than 99 % of our experiments. The biphasic activity of the enzyme correlates well with the lateral pressure signal of the expanding and condensing monolayer; thus, transient depletion is unlikely to be the origin of the observed cessation in activity. We wish to note that this biphasic behavior was consistently observed, even though the magnitude of the first phase of increased activity varies from measurement to measurement (for more examples, see **Fig. S2**). We further conclude from the conservation of the biphasic nature of the response, that the observed phenomenon is independent of the type of acid used to evoke excitation.

Higher time resolution is achieved, and more details can be resolved, when excitation is induced by $CO_2$ instead of by HCl. In water, $CO_2$ undergoes partial conversion to $HCO_3^-$ and $H_3O^+$ on a very slow timescale ($k_{H+} \sim 0.0375$ 1/s) [29]. Consequently, the resulting pulses are much broader than those generated by HCl (see Figs. 2C & D). Nevertheless, the biphasic activity pattern is conserved. During the expanding wave front activity increases 1.5-2.5-fold, while during the subsequent condensation back to equilibrium substrate cleavage ceases almost completely.

To confirm that we were indeed observing the impact of the propagating pulse on enzymatic activity, control experiments were performed under identical conditions (substrate, assay components, buffer, lipids), but in the absence of enzyme. No effect of the pulses on the substrate concentration were thus detected (Fig. 2E). Importantly, the quasi-adiabatic results obtained are in sharp contrast to the behavior of the enzyme during isothermal measurements: in this latter case the monolayer is quasi-statically compressed, resulting in an approximately constant catalytic rate in the particular pressure regime (Fig. S3) [30].

As demonstrated earlier [13,19], the pulses that we describe are acoustic, and should not be confused with the chemical waves in excitable materials that were reported earlier [31–33], in which chemical diffusion serves as the underlying mechanism for propagation. Here, no material transport is necessary for propagation and communication of a signal (Footnote: Since this initiates exploration of a new **concept in biology,** we use the word 'communication' in its broadest sense as "production of an effect at a distance by a localized event"). In addition, since both membrane leaflets can undergo the adiabatic change, communication between intra- and extracellular associated entities (e.g. proteins) can occur without any material transport across the membrane and does not require any molecular "transmitters" (*e.g.,* membrane-spanning proteins). Furthermore, in principle, no energy is required for transmission, given the adiabatic nature of the evoked pulse. Thermodynamically, it can be viewed as an (adiabatic) state change propagating along the interface. Consequently, an enzyme located within the acoustic path of the pulse experiences a transient change in all its thermodynamic variables. The extent of the state change will depend both on the amplitude of the propagating pulse and on the initial conditions within the membrane patch in which the enzyme is embedded. The fact that the lipid environment can regulate the activity of membrane-bound enzymes is well known from quasi-static experiments [34–39]. But in our system, whereas isothermal changes in the lipid monolayer in the liquid-expanded phase exert only a minor influence on enzyme activity (cf. S3B), the effect of the adiabatic pulses is striking.

As pointed out above, naturally all the observables of a system are coupled (cf. Maxwell relations) [40]. Thus, for example, when a charged membrane is compressed, not only the pressure and the temperature increase reversibly, but also the charge density and/or surface potentials. Hence, along with a pressure pulse a simultaneous propagating change in surface potential is anticipated, as indeed we have reported [15,19]. Consequently, ion distribution, in particular the distribution of the very mobile protons, will rearrange under the presence of the propagating electric field, resulting in the propagation of a pH pulse simultaneously with those in 2D-pressure, voltage and temperature, as we confirmed recently [19]. Reversible pH changes of up to +/-1 unit at the interface can be easily achieved. A pressure drop in our monolayer system is associated with a local increase in pH [19], which should indeed give rise to an increase in AChE activity. This increase arises most likely from the pH-dependence of AChE, which displays a pH-optimum at a bulk pH of ~7.5 [41]. A pH increase from 6.5 to 7.5, as observed during pulse propagation, should produce a ~2-fold change in activity, in excellent agreement with that achieved by $CO_2$ excitation (cf. Fig. 3). When, however, excitation is evoked by HCl, time scales shorten and activity changes increase. Although it is beyond the scope of this paper to discuss these results quantitatively, potential sources for this difference should be mentioned: a. Pulses are not isothermal but adiabatic, and activity-state relationships should be expected to vary in the same sense as all state diagrams depend on such conditions; Since HCl-excitation produces rapid local changes in the interfacial variables, the process resembles an ideal adiabatic state change much more closely than the relatively slow $CO_2$-induced changes, which fall somewhere in between the two extreme cases (adiabatic vs. isothermal). b. Taking the perspective of the protons - which are components of the reaction - the reaction equilibrium is shifted differently in the two cases. In the quasi-adiabatic case the rapid increase in pH due to the pulse results in a transient release of protons by the chemical reaction, which in turn creates a shift towards product production. This shift is much less pronounced in the quasi-isothermal case. Of course, interaction between pulses and enzymes is not limited to pH effects but is dependent on all the thermodynamic variables. Thus, for example, enzyme activity is often coupled to other nonlinear physical quantities, such as electrical capacity or compressibility, turning this general phenomenon into a very specific process [37,42,43].

One surprising observation is the asymmetric biphasic change in activity for the pH-induced pulse (cf. Fig. 2). Why does the time course of the change in enzyme activity not follow the time course of variation in pressure more faithfully? More specifically, why is the initial increase in activity followed by a phase in which the enzyme is inactive, rather than by a monotonic decrease back to the baseline activity level? As already established, every pulse represents a propagating thermodynamic state change. Hence its interaction with proteins will strongly depend on the initial state of the interface, the strength of the perturbation, and the physicochemical properties of the individual protein. We wish, therefore, to refrain from attempting to provide an explicit molecular description at this stage. Nevertheless, we hypothesize that the biphasic activity of the AChE in the case of the pH-induced pulses may be explained as follows: In the first phase (1) the pulse-induced pH increase leads to an increase in enzyme activity (Fig. 3). Due to the strongly enhanced degradation of substrate, the local proton concentration around the enzyme drastically increases. Combined with the back diffusion of the liberated protons from the lipids, this could lead to a temporary pH drop at the interface and hence to a decrease in enzyme activity (2) that would resemble the negative feedback observed for the pH-dependence of a native membrane-bound AChE preparation[30]. Finally, the system relaxes back to initial conditions as the perturbation disappears (3).

It would be informative and exciting to follow these catalytic changes at the single molecule level, which could be achieved by use of fluorescence correlation spectroscopy (FCS). Kaufmann has developed a theory based on the second law of thermodynamics, according to which changes in catalytic activity represent changes in fluctuation strength of the reaction coordinate induced by the absorption of the substrate[9,10,44]. In another manuscript we indeed demonstrate that the catalytic rate of AChE directly follows the state (compressibility), and hence the fluctuation strength of the interface in which the enzyme is embedded [30,39]. We intend to further test this theory by coupling our experimental system with FCS technology to access single molecule fluctuations similar to those shown in the work of Rigler and coworkers[45].

In conclusion, we wish to mention that we performed a series of experiments with the enzyme phospholipase A2. Initial results show that the enzymic activity decreases during enzyme pulse-interaction. We propose, therefore, that acoustic pulses provide a general mechanism for orchestrating intracellular processes (Fig. 4).

## Methods

**Chemicals**:

Acetylthiocholine (ATC), Ellman's reagent (5,5'-dithiobis-[2-nitrobenzoic acid]) (DTNB) and hydrochloric acid (32%) were purchased from Sigma-Aldrich (St. Louis, MO). 1,2-dimyristoyl-*sn*-glycero-3-phospho-L-serine (DMPS) was acquired from Avanti Polar Lipids (Alabaster, AL) and used without further purification. It was dissolved in a chloroform/methanol/water **(65/25/4 vol%)** solution at a concentration of 10 mg/ml. Detergent-soluble acetylcholinesterase from electric organ tissue of *Torpedo californica* was purified by affinity chromatography after solubilization in 1% sodium cholate[24].

**Experimental setup:**

For the enzyme-pulse experiments a custom-made Langmuir trough (NIMA, UK) with an area of 150 cm$^2$ (Fig. S1) was used. The propagating mechanical component of the pulse was measured by a pressure sensor with up to 10 000 samples per second and 0.01 mN/m resolution. The sensor was situated next to the window of detection (cf. Fig. S1), through which AChE activity was monitored using the chromophoric Ellman assay. The hydrolysis of ATC leads to the release of thiocholine, which immediately reduces DTNB, yielding an intense yellow chromophore with an absorption peak at ~412 nm [27,28]. Changes in the absorbance of the light emitted by a LED located above the trough, due to generation of the chromophore, were followed using a photodiode placed underneath the window (cf. Fig. 1).

**Experimental procedure:**

In order to study the effect of the HCl-induced pulses on enzyme activity the trough was filled with ultrapure water (resistivity >18 MΩcm), containing 100 mM NaCl, 10 mM phosphate, 2 mM ATC and 2 mM DTNB (initial pH 7.0). For the experiments that utilized $CO_2$ excitation the composition was 10 mM NaCl, 1 mM phosphate, 2 mM ATC and 2 mM DTNB (initial pH 7.0). A DMPS monolayer is generated by titrating a few microliters of the phospholipid solution onto the surface of the aqueous solution until the desired lateral pressure is obtained. After 10 min solvent evaporation, 3 μl of the enzyme solution (concentration 68 μg/ml) are added in the left part of the trough. To ensure homogeneous enzyme distribution the subphase is gently stirred for 5 min. After a further 15 min equilibration experiments are initiated. HCl-induced pulses are evoked by blowing 25 ml of $N_2$ through the gas phase of a bottle filled with 32% aqueous HCl. For the $CO_2$-induced pulses, pure $CO_2$ gas is blown onto the lipid monolayer for 15 s at a flux of 150 ml/min.

**Acknowledgments**

M.F.S. especially thanks Dr. Konrad Kaufmann (Göttingen), whose contributions to this work were crucial. He not only initiated this project, but also inspired him to work in this field. We also thank him for numerous seminars and discussions. We are very grateful to Prof. Achim Wixforth and his chair (Experimental Physics 1 – University of Augsburg) for the support of this project. M.F.S. appreciates funds for a guest professorship from the Deutsche Forschungsgemeinschaft (DFG), SHENC-research unit FOR 1543. B. F. acknowledges funding by a Studienstiftung des deutschen Volkes and DFG (FOR 1543).

## Author Contributions

M.F.S. and B.F. designed the experiments. B.F. performed the experiments. I.S. purified the AChE. All authors contributed to the preparation of the manuscript.

## Competing financial interests

The authors declare no competing financial interests.

## Materials & Correspondence

Correspondence should be addressed to M.F.S. (matthias-f. schneider@tu-dortmund.de).

## Figures

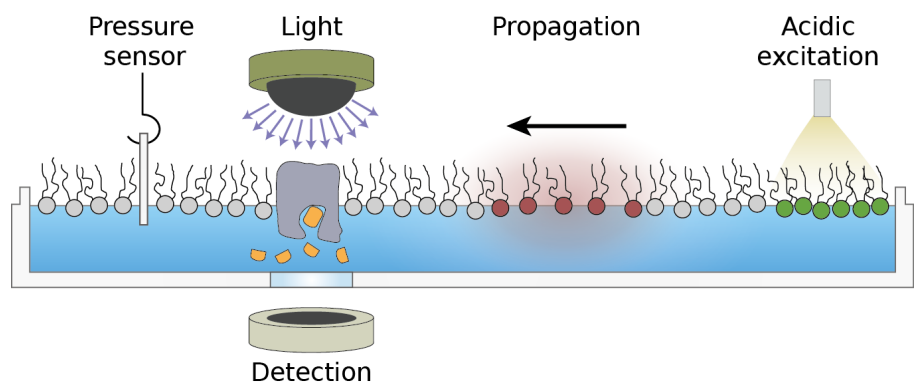

**Fig. 1**. Langmuir trough setup for measuring pulse-enzyme interaction: pulses are excited in a lipid monolayer via local application of gaseous acid (HCl or $CO_2$). The protonation of the lipid head groups evokes a propagating pulse whose mechanical component is recorded by a pressure sensor. Simultaneously, the activity of the enzyme acetylcholinesterase is monitored by the colorimetric Ellman assay, which is based on the absorption of the reaction product at ~ 410 nm.

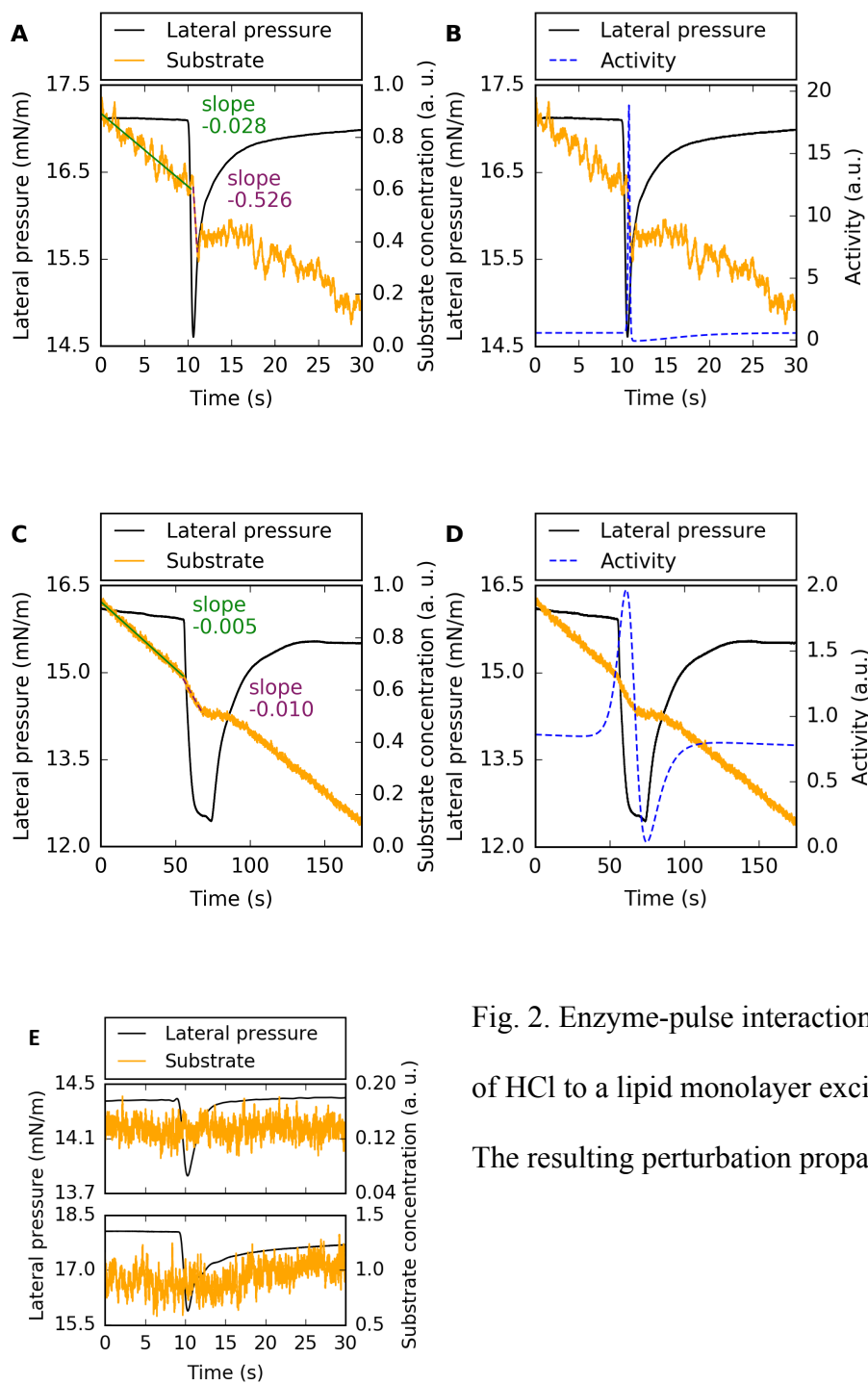

Fig. 2. Enzyme-pulse interaction: (A) The local application of HCl to a lipid monolayer excites a lateral pressure pulse. The resulting perturbation propagates with a velocity of

~0.5 m/s across the monolayer. The enzyme AChE is incorporated into the interface, and responds to the pulse by a change in its catalytic rate, which is visualized by the nonlinear change in substrate breakdown. The slope of the curve corresponds to the activity of the enzyme, which increases ~19-fold during the first (expanding) phase of the pulse (slope 1 = -0.028, slope 2 = -0.526, factor = 18.8). During the subsequent condensation back to equilibrium substrate cleavage stops completely for ~5 s before returning to a value similar to that which preceded the pulse. The activity change is illustrated by the blue dashed line in (B), and represents a guide to the eye. The ambient buffer composition is 100 mM NaCl/10 mM phosphate, pH 6.5, at 25°C; (C) $CO_2$ can also be used to evoke lateral pressure pulses. The slow conversion of $CO_2$ to $HCO_3^-$ and $H_3O^+$ in water results in very long wavelengths, and hence in more precise intensity measurements. In this case, too, pulse-enzyme interaction manifests as a biphasic change in activity, increasing during the expanding phase (~2-fold) and ceasing transiently during the subsequent condensing phase. Panel (D) depicts another guide to the eye in order to illustrate the biphasic activity change. (E) Controls, measured in the absence of AChE. The substrate concentration is not affected by the pulse. The ambient buffer composition is 10 mM NaCl/1 mM phosphate, pH 6.5, at 25°C.

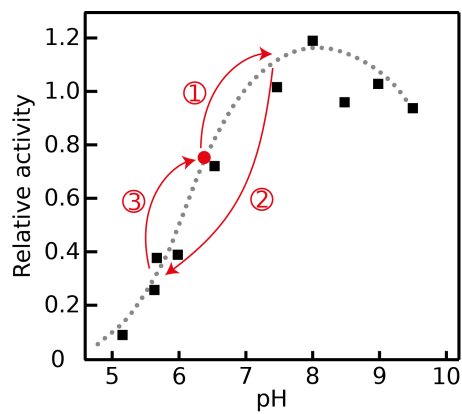

**Fig. 3.** pH-dependence of AChE activity (adapted from [41]): Measurements at equilibrium are performed at pH 6.5, indicated by the red dot. The propagating pulse reversibly alters the pH at the interface. Due to the pulse-induced local pH increase at the interface, increased enzymatic activity is expected initially (1). Consequently, the product concentration around the enzyme rises, and hence also the proton concentration. The protons released during the initial expanding phase migrate back to the surface during the subsequent condensation of the interface to equilibrium. This could produce a local proton excess, thus lowering the enzyme's catalytic rate (2). In the final step, the interface reverts to its initial state, and both the local pH and the enzymatic activity return to their initial equilibrium values (3).

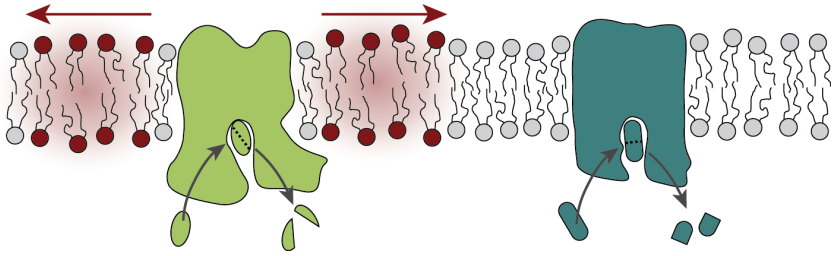

**Fig. 4.** Biological signaling by sound pulses: a perturbation, in this case induced by an enzymatic reaction, evokes a propagating signal across the interface. The excitation process may be very specific, since it depends on the initial state of the membrane as well as on the excitation strength. This specificity is, for instance, demonstrated by pH excitation of lipid monolayers, which depends directly on the p$K_a$ values of the constituent lipids [19]. A second enzyme, situated in the path of the travelling pulse, will experience a transient change in all its variables. If the pulse-induced changes are strong enough, its activity will be temporarily modulated. Hence interfacial sound pulses provide a plausible mechanism for specific signaling within cells.